# Robert Grosseteste's thought on Light and Form of the World

## Amelia Carolina Sparavigna

Department of Applied Science and Technology, Politecnico di Torino, Torino, Italy

**Abstract:** Robert Grosseteste was one of the most prominent thinkers of the Thirteenth Century. Philosopher and scientist, he proposed a metaphysics based on the propagation of light. In this framework, he gave a cosmology too. Here we will discuss the treatise where Grosseteste proposed it, that entitled "De luce, seu de incohatione formarum", "On Light and the Beginning of Forms".

## 1. Introduction

Robert Grosseteste, Bishop of Lincoln from 1235 to 1253, was one of the most prominent and remarkable thinkers of the Thirteenth Century. He was philosopher and scientist. As told in [1], he was heavily influenced by Augustine, whose thought permeates his writings and from whom he drew a Neoplatonic outlook. However, as told in [2], he also made extensive use of the thought of Aristotle, Avicenna and Averroes.

Grosseteste lived in a period during which two main factors dominated the culture. One was the birth in the Western Europe of universities, which can be regarded as the evolution of previous modes of instruction; the second factor was the impact of Islamic philosophy, which was the vehicle to the West of the integral knowledge of Aristotle [1]. In this framework, Grosseteste developed an original account of the generation and nature of the physical world in terms of the action of light. The world is the Latin "mundus", which is Earth and heavens, and in the Grosseteste's thought had the form of spheres as in the ancient Aristotelian cosmos.

Besides philosophical treatises, Grosseteste composed several short works regarding optics and other natural phenomena, such as sound and heat, which are quite interesting, and of which we have proposed translations and discussions [3-8]. Here we will read one of his treatises, that entitled "De luce, seu de incohatione formarum", "On Light and the Beginning of Forms", where Grosseteste is proposing his metaphysics, and a consequent cosmology based, on light. However, first of all, let us see the discussions of some scholars on this Grosseteste's treatise.

## 2. The light metaphysics

The item on Grosseteste in the Stanford Encyclopedia of Philosophy is written by Neil Lewis of Georgetown University [2]. Lewis defines light as the leitmotif running through Grosseteste's works. Besides in his writings on optics, light occupies a prominent place in all Grosseteste's works, in accounting of sense perception which is relating body and soul, in his theory of knowledge based on illumination, and in the origin and nature of the physical world [2].

Grosseteste's metaphysics rests on a hylomorphic account of the nature of bodies (the hylomorphism is a philosophical theory developed by Aristotle, which conceives being as a compound of matter and form); according to this approach, the philosopher proposed his cosmology, described in the "De Luce". We find that the firmament, which is the outermost heavenly sphere, is the simplest body of the world, composed of first matter and first form. To Grosseteste, the first form was light (lux).

The De Luce opens with an argument for the identification of first form with lux: first form and first matter are in themselves simple substances. The first form, which is also called "corporeity", is coming from the extension of matter into three dimensions, thereby yielding a dimensioned body [2,9,10]. An entity without dimension could only have this effect if it instantaneously multiplied and diffused isotropically in all directions. In fact, these are features of light, for light is essentially self-multiplicative and self-diffusive, and a sphere of light being instantaneously generated from a point source. Moreover, light is dragging matter along with its diffusion. Thus, Grosseteste concluded that light is in fact the first form.

He saw in the "light metaphysics" some of God's creation of the physical universe and an explanation of why it took its form [2]. At the beginning of time, God created first form, lux, in first matter. As explained in [2], first form and first matter are in themselves indivisible and simple. And, according to Grosseteste, the finite multiplication of a simple cannot generate an item with size. But, the infinite multiplication of a simple will generate a finite quantity (quantum). Therefore, through the infinite multiplication of first form in first matter, extended bodies can be produced and thus the physical universe. It is remarkable that, to account for bodies of different sizes, Grosseteste argued that there are infinities of different sizes that stand in different ratios [2].

Let us add to Lewis' observations, which are like those we can find in the discussion and translation of De Luce given by Clare C. Riedl [11], some other comments. Grosseteste's first form indivisible and simple, in Latin "simplex", is similar to the indivisible of Bonaventura Cavalieri's calculus [12]. These indivisibles were dimensionless until the development of differential calculus by Newton and Leibniz. The sum of these indivisibles is able to give lines, surfaces and volumes. In the translation of De Luce proposed in this paper, I consider the Grosseteste's indivisible as an entity, that is, something that exists in itself, referring to it as a dimensionless being. In [11], it is used the term "being".

Grosseteste used the light to explain the genesis of the Aristotelian cosmos as a system of nested celestial spheres surrounding the four sublunary or elemental spheres. Matter and light existed at the beginning of time at a single point. The infinite self-multiplication of the initial point of light extended the first matter into a spherical form, because light diffuses itself spherically. Let us observe, that what Grosseteste tells is a manifest anticipation of Huygens' theory of the propagation of light [7,13]. The outermost parts of the matter of the sphere thereby generated, unlike the parts of the matter below them, were maximally extended and rarefied and formed the outermost sphere, or first sphere, the firmament [2]. Since the light is essentially self-multiplicative, this light in this outermost sphere continued to multiply itself, but back inwards, toward the center from all parts of the outermost sphere, because it had already diffused itself outward as far as it possibly could. However, the light, "being a substantial form, cannot exist apart from matter, this inwardly directed light drew with itself what Grosseteste calls the spirituality of the matter of the outermost sphere, and thus luminosity (lumen), a body comprised of light and the spirituality of this matter, proceeded inwards. Moving inwards, luminosity concentrated the matter existing below the outermost sphere, leaving in its wake below the outermost sphere a second sphere comprised of matter whose parts were rarefied as much as they could be. This sphere in turn generated luminosity, which moving inwards further concentrated the matter below it and rarefied the outermost parts of this matter so as to produce the third sphere. The repetition of this process gave rise to the nine celestial spheres, each sphere being comprised of matter whose parts were incapable of further rarefaction" [2].

This description proposed by Lewis, and in particular his referring to a wake, that is, a visible track of turbulence left by something moving through water or air, allows us to imagine the Grosseteste's universe as formed by spherical waves in the volume of the universe, as the concentric waves on the surface of water. Grosseteste knew this phenomenon for sure and then he used it for modelling the Aristotelian world.

Let us continue the Grosseteste's description of world creation. The lowest celestial sphere, the lunar sphere, also generated luminosity, which moved inwards and concentrated the matter below it. But this luminosity had a low power and therefore produced a sphere comprised of incompletely dispersed matter, the sphere of Fire. Likewise, fire generated luminosity producing below it the sphere of Air. This process continued, giving the spheres of Water and Earth, the latter being comprised of the most concentrated and dense matter. These classical four elements, unlike the celestial spheres, are capable of alteration, growth, generation and corruption [2].

For what concerns the motion of the heavenly bodies, Grosseteste tells they can only move with a circular movement because the luminosity in them is incapable of rarefaction or condensation, and as a result cannot incline the parts of their matter upward, to rarefy them, or downward, to condense them [2]. Then, the heavenly spheres receive movement from an intellective motive power. But the elements can be rarefied and condensed; they can incline the luminosity in themselves away from the center of the universe, so as to rarefy it, or toward the center so as to condense it, and this accounts for their natural motion up and down [2].

The metaphysics of light is an original idea that Grosseteste used to explain the world as imagined by Aristotle, a machine consisting of nested spheres, and the distinction between the motion of celestial and sublunary bodies. In the Figure 1, we can see these spheres, depicted in an Icelandic manuscript. It is a geocentric world, with the firmament above the sphere of the Zodiac.

**3. Grosseteste and the Augustinian thought**

Ralph Matthew McInerny (1929 – 2010) was a Professor of Philosophy at the University of Notre Dame. He wrote about Grosseteste in the second volume of "A History of Western Philosophy", in the chapter on the philosophy of the Thirteenth Century. This chapter starts with the discussions on thoughts of William of Auvergne and Alexander of Hales, at the University of Paris. The works of these philosophers are good examples of the new philosophical writings that had started in the Western universities, during that century. Meanwhile, McInerny continues, "at Oxford the example of Robert Grosseteste is an indication of a quite different response to the new literature. Robert, who was later to become bishop of Lincoln, was well acquainted with the works of Aristotle. … The thing that strikes the reader of the philosophical writings of Grosseteste, edited in 1912 by Ludwig Baur, is the preponderance of mathematical and scientific topics. It is easy to feel that here is independence and originality of a sort unknown in William of Auvergne and Alexander of Hales. Further consideration leads, however, to the judgment that, despite the mathematics, Grosseteste is actually representative of a conservative mentality, that in him Augustinianism lives on in a less adulterated form than in his continental contemporaries. It is customary, convenient, and fitting that the flavor of Grosseteste's work be exhibited by his contribution to Augustine's theory of illumination" [1].

Then, McInerny continues with the discussion of the De Luce. "The following amounts to a rough translation of the beginning of that essay. I think, Grosseteste writes, that the first bodily form, what some call corporeity, is light, for light of its very nature (per se) diffuses itself in all directions such that, given a point of light, a sphere of light of whatever size is immediately generated unless something opaque (umbrosum) impedes. Matter's extension in three dimensions follows necessarily on corporeity, but matter itself is a simple substance lacking dimensions. So too, form is a simple substance also lacking dimensions, and it cannot account for the dimensions matter comes to have. To account for the extension of matter, Grosseteste says, I nominate light. Extension in all directions is a per se property of light; it diffuses and multiplies itself everywhere. Whatever performs the task of introducing dimensions into the compound of form and matter must therefore be either light or something that does this just insofar as it participates in light. Corporeity, bodily extension, is either light or a participation in light: something which acts through the power of light. Grosseteste's own opinion is simply put. Light is the most noble form of bodies and is that in bodies which makes them most akin to separate substances" [1].

Grosseteste used light to explain the extension of bodies, therefore he used it to explain the constitution of the universe too. "We mentioned earlier that the diffusion of light can be checked by the interposition of an obstacle; Grosseteste also holds that any given point of light has an intrinsic limitation on the extent of its diffusion. As for the constitution of the cosmos, then, he can begin with a single body which may be thought of as light and matter, a compound of form and matter: its diffusion to the extent of its intrinsic power will produce a sphere which is finite and whose limit is the heaven. Then, by thinking of that outer limit of light reflecting on the center from which it radiated, Grosseteste speaks of the generation of the celestial bodies. The picture that results is quite geocentric. The degree or intensity of light provides Grosseteste with a scale on which he can compute the ontological status of entities, so that the universe for him is a hierarchy of lights or a hierarchy based on degrees of participation in light. Thus far Grosseteste's use of light to explain the cosmos may seem only the inspiration of one who had been impressed by the application of mathematics to natural phenomena, like the distribution of light from a source and like the rainbow. … At any rate, beyond his attempt to interpret the physical world by means of light as his basic concept, Grosseteste's theory must be seen as a continuation of the Augustinian doctrine of illumination. St. James spoke of God as the Father of lights and St. John of Christ as the light of the world, and it may not be too much to say that what Augustine had developed from such scriptural remarks as these is as important for the development of Grosseteste's universe of light as anything of an observational nature" [1]. To his discussion, Ralph McInerny suggested references [14-17].

**4. A remarkable man of science**
Clare C. Riedl proposed in 1942 a translation and discussion of De Luce [11]. This translation is the one commonly used for studying the Grosseteste's treatise. In [11], it is told that Grosseteste was "without question one of the most remarkable men of science of his time". In fact, he studied optics, in particular the reflection and refraction laws, pointing the way to microscopes and telescopes. In philosophy, "Grosseteste represents, and indeed might be called the founder of, a new tradition, characterized by the blending of philosophy with experimental science. This tradition continued to be characteristic of philosophy at Oxford in opposition to the more metaphysical type of speculation which prevailed at Paris" [11].

According to Riedl, De Luce is significant because is an example of the philosophic-scientific synthesis of the Oxford School, and it was an important source of the "light metaphysics" and was fundamental for the medieval conceptions relative to cosmology. To understand the treatise, it is necessary to consider some aspects of Grosseteste's doctrine of matter and form: the terminology is Aristotelian but ideas are original. According to Grosseteste, matter is not pure potency, as it was for Aristotle, but possesses in its own right a certain minimal reality. Matter is a substance then, and the form completes, perfects and actualizes it, giving it a dimension.

The first corporeal form is the light: it is more than the form of corporeity, it is also a principle of activity. "Every body, he (Grosseteste) believes, has a motion or activity which is natural to it, because it proceeds from an intrinsic principle. The intrinsic principle from which this motion or activity proceeds must be the form, since matter is passive" [11].

The De Luce can be considered as composed of two parts, the first is concerned with "light metaphysics" proper, the second contains a cosmogony obtained from this metaphysics. Grosseteste bases his theory on the fact that a characteristic note of corporeity is the requirement of extension in the three dimensions. He knows the property of light to diffuse in all directions, multiplying itself, and that a point source is producing a sphere. Therefore, he concludes that the light is suited to fulfil the requirement of extension and when it is joined to matter as its form, being inseparable from it, the light will necessary carry matter along with it in its diffusion and self-multiplication. In fact, the light of which Grosseteste speaks is not the ordinary physical light but a simple substance, almost spiritual in its properties. Moreover, Grosseteste uses two words: lux and lumen. The first form is the lux, light, whereas the reflected or radiated light is the lumen, that is the luminosity [11].

In the second part of the treatise, Grosseteste is proposing a philosophy of the Genesis. Riedl remarks that in this philosophy, it is the light which is giving the principle of continuity in nature, for, as being the first corporeal form, it is common to all things in the universe from the lowest of the elements, earth, up to and including even the firmament. The universe (mundus) with its thirteen spheres is the typical medieval world. It is geocentric with the ninth heavenly sphere that Ptolemy added to Aristotle's eight. Moreover, it seems that the cosmology of the De Luce shows considerable traces of the influence of Alpetragius (Al-Bitrogi) [11].

On the movements of the spheres, Grosseteste tells that there is the diurnal motion. It is imparted by the outermost sphere, the firmament. This is a somewhat new theory in Grosseteste's day, the suggestion coming from an Arabian writer, Thebit ben Coran (Ibn-Thabit), referred frequently by Grosseteste [11].

It's time to read the treatise, of which I am proposing a translation using some physics and mathematics terms. The original Latin text is freely available at www.grosseteste.com.

**5. Grosseteste's On Light**
The first corporeal form, which is also referred to as "corporeity", is in my opinion the light, because the light, lux in Latin, due to its very nature, diffuses itself in every direction in such a way that a point source will give instantaneously a sphere of light of any size, unless some object producing shadows is obstructing its rays. Corporeity is coming from the extension of matter in the three dimensional space, and this happens in spite of the fact that both corporeity and matter are in themselves substances lacking of dimension. But a form, which is in itself simple and dimensionless could not induce dimensions in every direction into the matter, which is likewise simple and dimensionless, except by multiplying itself and diffusing itself instantaneously in every direction and thus extending matter in its own diffusion. It is so, because the form cannot abandon matter, since it is inseparable from it, and matter itself cannot be deprived of form. Now, let us consider light, which has its nature characterized by the property of being able of multiplying itself and diffusing itself instantaneously in all directions. Whatever is acting, either light or a participation in light, that is something which acts through the power of light, we have an agent which accomplishes this operation by itself. Corporeity, therefore, is either the light itself or an agent which performs the operation previously mentioned and is able to induce dimensions into matter, as a result of participating in light, and acting through the power of it. But the first form cannot induce dimensions into matter through the power of a consequent form. Therefore, light is not a form consequent to corporeity, but is corporeity itself.

Moreover, it is opinion of scholars that the first corporeal form is the worthier and nobler and more excellent essence than all the forms coming after it. It has a high resemblance to the forms that are existing separated (from matter). That is, light is the worthier, nobler and more excellent essence than all corporeal things. It is more than all other bodies similar to the forms that exist separated (from matter), namely, the intelligences. Light therefore is the first corporeal form.

Due to its nature, light, which was the first form created in first created matter, multiplied itself an infinite number of times and expanded itself isotropically in all directions. In this way, to the very beginning of time, light caused the spreading of matter, that could not leave behind, by pulling it along with itself, into a quantity equal to the mass of the entire machine of the world [18]. Let us stress that this extension of matter could not be obtained through a finite multiplication of light, because the multiplication of a simple entity a finite number of times does not produce a "quantum", that is a quantity, as Aristotle shows in his De Caelo et Mundo. However, the multiplication of a simple entity an infinite number of times must give a finite quantity, because a product which is the result of an infinite multiplication exceeds infinitely the entity multiplied. Now, one simple entity cannot exceed another simple entity infinitely: only a finite quantity infinitely exceeds a simple entity. Therefore, an infinite quantity exceeds a simple entity by infinity times infinity. When light, which is in itself an entity, is multiplied an infinite number of times, it must extend matter, which is likewise an entity, into finite dimension.

It is possible, however, that an infinite series of terms is related to an infinite sum in every proportion, numerical and non-numerical [19]. Some infinites are larger than other infinites, and some are smaller. Thus, the series of all numbers, even and odd together, is infinite. At the same time, this series is greater than the series obtaining from all the even numbers, which is infinite too, because it is exceeding it by the series of all the odd numbers. The series, too, of all numbers starting with one and continuing by doubling each successive number is infinite, and similarly the series of all the halves of these doubles is infinite. The series of halves must be half of the series of doubles. In the same way the series obtained from all numbers starting with one and multiplying by three successively is three times the series of thirds corresponding to them. It is likewise clear, for all given numerical proportions, that we can have a proportion of finite to infinite according to each of them.

Now, let us consider an infinite series of all doubles starting from one, and an infinite series of all the halves corresponding to these doubles: if one, or some other finite number, is subtracted from the series of the halves, after this subtraction we will have no longer a two to one proportion between the first series and what is left of the second series. Therefore, there will not be any numerical proportion. The reason is the following: "if a second numerical proportion is to be left from the first as the result of subtraction from the lesser member of the proportion, then what is subtracted must be an aliquot part or aliquot parts of an aliquot part of that from which it is subtracted. But, a finite number cannot be an aliquot part or aliquot parts of an aliquot part of an infinite number [20]" . Therefore when we subtract a number from an infinite series of halves, we have a non-numerical proportion between the series of doubles and what is left from the series of halves. (So we have numerical and non-numerical proportions).

After this discussion, it is clear that light through the infinite multiplication of itself extends matter into finite dimensions, that can be smaller and larger according to certain respective proportions, numerical and non-numerical. In fact, if light through its infinite multiplication extends matter into a dimension of two cubits, by the doubling of this infinite multiplication, it extends it into a dimension of four cubits, and by the dividing in half, it extends it into a dimension of one cubit. Thus it proceeds according to numerical and non-numerical proportions.

In my opinion, this was the meaning of the theory of those philosophers who told that everything is composed of atoms, and that bodies are composed of surfaces, surfaces of lines, and lines of points. This opinion does not contradict the theory that a magnitude is composed only of magnitudes. In fact, "whole" is said in so many ways as "part" is said. Thus, we say that a half is part of a whole, because two halves make a whole. We say that a side is part of a diameter, but this is said in a different meaning: no matter how many times a side is taken, it does not make a diameter, but it is always less than the diameter [21]. Again, we say that an angle of contingence is part of a right angle because there is an infinite number of angles of contingence in a right angle. When an angle of contingence is subtracted from a right angle a finite number of times the latter becomes smaller [22]. Differently, a point is a part of a line in which it is contained an infinite number of times; when a point is removed from the line a finite number of times this does not shorten the line.

To return to the main subject of this treatise, I say that light through its isotropic infinite multiplication extends matter into the form of a sphere and, as a necessary consequence of this extension, the outermost parts of matter are more extended and more rarefied than those inside the volume, close to the center of the sphere. Since the outermost parts of the sphere become highly rarefied, the inner parts have the possibility of further rarefaction.

In such a manner the light acted, by extending first matter forming a sphere and by rarefying its outermost parts to the highest possible degree. In this outermost part of the sphere, light fully completed the potentiality of matter, and left this matter without any susceptibility of further impression. Therefore, the first body in the outermost part of the sphere of the world, which is called "firmament", is perfect, because it has nothing in its composition but first matter and first form. It is therefore the simplest of all the structures of the world, having the greatest possible extent, with respect to the parts that constitute its essence and with respect to its quantity.

Firmament is an object of the category "body", with the specific property that, in it, the matter has a complete actuation through the first form alone. But the "body", which is in this and in other bodies, has in its essence first matter and first form, and so it drives the matter to complete it through the first form and to reduce it through the first form.

When the first structure, the firmament, was complete, it diffused its luminosity, the Latin lumen, from every part of itself towards its center. Lux, the light, after the fulfilment of the first body, naturally multiplied itself from it, and necessarily diffused to its center. And since lux is form inseparable from matter, during its diffusion from the first body to the center, extended along with itself the spirituality of the first body matter. And thus, we have a diffused light, a luminosity, coming from the first body, and this luminosity is a spiritual body, or if you prefer, a bodily spirit. This luminosity, in its transit, does not split the medium through which it is passing, and thus it passes instantaneously from the bulk of the first heaven to the center of the sphere. Furthermore, its passing is not to be understood in the sense of an entity passing instantaneously from that heaven to the center, for this is perhaps impossible, but its passing takes place through the multiplication of itself and an infinite generation.

This luminosity, expanded from the first body toward the center and gathered together the mass existing below the first body; and since the first body could no longer be lessened because it was completed perfectly and invariably, and since, too, there could not be an empty space, it was necessary, in this gathering of mass, the outermost parts be disgregated and expanded. Therefore, the inner parts of the aforesaid mass became denser and the outer parts rarefied. So great was the power of this luminosity, which was gathering together mass, aggregating and disgregating matter, that the outermost parts of the mass contained below the first body were elongate and rarefied to the highest degree. Thus, in the outermost parts of the mass in question, the second sphere was created, completed and susceptible of no further impression. And this is the creation and perfection of the second sphere, where we have luminosity generated by the first sphere and the light, which is simple in the first sphere, is doubled in the second.

Just as the luminosity generated from the first body completed the second sphere, leaving a denser mass below the second sphere, so the luminosity generated from the second sphere completed the third sphere, leaving below this third sphere a mass of even greater density, after aggregation and disgregation. This process of simultaneously aggregation and disgregation continued in this way until the nine heavenly spheres were completed, gathering together, below the ninth and lowest sphere, the dense mass which constitutes the matter of the four elements.

The lowest sphere, the sphere of the Moon, which is generating luminosity from itself too, by its luminosity aggregated the mass contained below itself and, after this aggregation, rarefied and expanded its outermost parts. However, the power of this luminosity was not so great to produce a further expansion of the outermost parts of this mass to the highest degree. For this reason, mass was left imperfect and capable of being aggregated and disgregated.

The highest part of this mass was disgregated, although not to the greatest possible extent. By its disgregation, fire is coming and the matter of elements remains. This fire, generating luminosity from itself, aggregated the mass contained below it, with the disgregation of its outermost parts, but not completely, and in this way it produced air. Air, also, generated from its spiritual body or from its bodily spirit, produced water and earth by means of aggregation of inner parts and disgregation of its outer parts. But because water retained more of the power of aggregation than of disgregation, water remained together with the heavy earth.

In this way, therefore, the thirteen spheres of our world were created. Nine of them, the heavenly spheres, are not subject to change, increase, generation or corruption because they are complete, that is, perfect. The other four spheres have the opposite mode of being, that is, they are subject to change, increase, generation and corruption, because they are incomplete. It is clear that every higher body, in virtue of luminosity which proceeds from it, is that body featuring the body that comes after it. And like the power of unity is in every number that comes after it, so the first body, through the multiplication of its luminosity, is in every body that comes after it.

Earth is from the aggregation in itself of higher luminosities from all the higher bodies. For this reason earth is called Pan by poets, that is 'the whole,' and it is also given the name Cybele, which is like "cubile", from cube, that is, a solid. For this reason earth, that is, Cybele, which is the most compact of all bodies, is the mother of all the gods; because in her the higher lights are gathered together, however not driven for her own operation, but the luminosity of any sphere can be raised from it into act and operation. Thus every one of the gods can be considered generated from her as from of a mother. The intermediate bodies have a twofold behaviour. Towards lower bodies they have the same behaviour as the first heaven to all remaining things, and, they are related to the higher bodies as earth is related to all further things. And in this manner, some features of them remain in everything.

The image and perfection of all bodies is light, but in the higher bodies it is more spiritual and simple, whereas in the lower bodies it is more corporeal and multiplied. Furthermore, all bodies have not the same features, even though they all proceed from light, whether simple or multiplied, like the numbers, which are not all of the same kind, in spite of the fact that they are all derived from unity by a greater or lesser multiplication. And in this discussion, perhaps, we find the meaning of the sentences telling that "all is one, in the perfection of one light" and also, "those, which are plural, are plural through different multiplication of light itself".

Since lower bodies participate in the form of the higher bodies, a lower body because of its participation in the form of the higher body, receives its motion from the same incorporeal moving power by which the higher body is moved. Therefore, the incorporeal power from an intelligence or spirit, who moves the first and highest sphere in the diurnal motion, moves all the lower heavenly spheres in the same diurnal motion. However, these spheres receive their motion weakened in proportion as these spheres are lower, because purity and strength of the first corporeal light in it is proportionally lower.

So we have the elements participating in the form of the first heaven. However, they are not moved by the mover of the first heaven in a diurnal motion. Although they participate in that first light, they are not obedient to the first moving power, since that light in them is impure, weak, and the purity, which it has in the first body, diluted; moreover, they possess the density of matter which is the principle of resistance and disobedience. However, there are some who believe that the sphere of fire rotates with a diurnal motion, and they consider the rotation of comets to be a sign of this. They say also that this motion is available in the waters of the sea, so that the tide of the seas is coming from it. But the right philosophers say that the earth is free from this motion.

In this same way, too, the spheres that come after the second sphere, the Zodiac, usually called the eighth when we call them from the earth upwards, all are transmitting the motion of this sphere, because they participate in its form. Therefore, this motion is proper to each of them in addition to the diurnal motion.

As previously told, the heavenly spheres are perfect and are not receptive of rarefaction or condensation, the light in them does not strain the parts of matter either away from the center, to rarefy them, or toward the center to condense them. And then, the heavenly spheres are not receptive of up or down motion but only of circular motion due to an intellectual moving power, which, by looking at itself in a corporeal manner, revolves the spheres to have a circular corporeal motion. On the contrary, elements are incomplete, and then subjected to rarefaction and condensation; the luminosity which is in them inclines them away from the center so as to rarefy them, or toward the center so as to condense them. And on this account they are naturally capable of being moved in an upward or downward motion.

In the highest body, which is the simplest of all bodies, there are four features, namely form, matter, composition and composite. The form, being the simplest, holds the position of unity. Matter, because of twofold potency, namely its susceptibility to impressions and its receptiveness of them, and also for its density which is proper to matter, which is primarily and principally a characteristic of duality, rightly selects a dual nature. But composition has a trinity in itself because there appears in it matter with form and form with matter, and the typical property of the composition, which is found in every composite as a third feature, distinct from matter and form. And we have also the composite proper, after these three constituents, which is considered as a quaternity. Therefore, in the first body, in which all other bodies exist virtually, there is a quaternity and therefore the number of the remaining bodies is basically no more than ten. Because we have one coming from the form, two from matter, three with the composition and four from the composite: when they are added make a total of ten. Therefore ten is the number of the spherical bodies of the world, because the sphere of the elements, although it is divided into four, is nevertheless one by its participation in earthly corruptible nature.

From these considerations it is clear that ten is the universal perfect number, because every perfect whole has inside something like form and then it is a unity, and something like matter and then it is duality, something like composition and then it is a trinity, and something like composite and then becomes quaternity. We cannot add a fifth to these four. For this reason, every perfect whole is ten. It is clear also that only five proportions, found in these four numbers, one, two, three, four, are suitable for composition and to have the harmony able to stabilize every composite. And only these five proportions are the harmonies we find in musical melodies, in pantomimes and rhythmic measures. This is the end of the treatise on light of the Lincolnian.

**6. Conclusion**

In preparing this translation, I used several terms that I have already used in writing the previous papers [3-8], for instance, the "disgregation", which is in Reference 8 too, a reference discussing the four classical elements, fire, air, water and earth. Here we have seen the light, the pivot about which the nature is turning.

Due to its importance, the De Luce had been discussed by several scholars; of some of them we have already reported their conclusions. We have also shortly discussed it in [23], a reference considering the form of the world during the Middle Ages. The De Luce contains several ideas suitable to a comparison to modern physics and mathematics. As told by J. Cunningham in Ref.24, in this treatise we find that Grosseteste imagined that matter existed at the beginning of time at a single point, "except that it did not exist in any sense that we would understand since it had no dimensions. It was neither three, two or even one-dimensional. It was therefore innate, but without existence in either time or space. Then God issued His first fiat saying 'Let there be light' and at that instant a shared of light issued from the divine and entered matter. As it did so it gave to matter dimension; or as Grosseteste would put it, 'corporeity.' This was 'the beginning' … No-one can read the Tractatus de luce today without thinking of the Big Bang theory" [24].

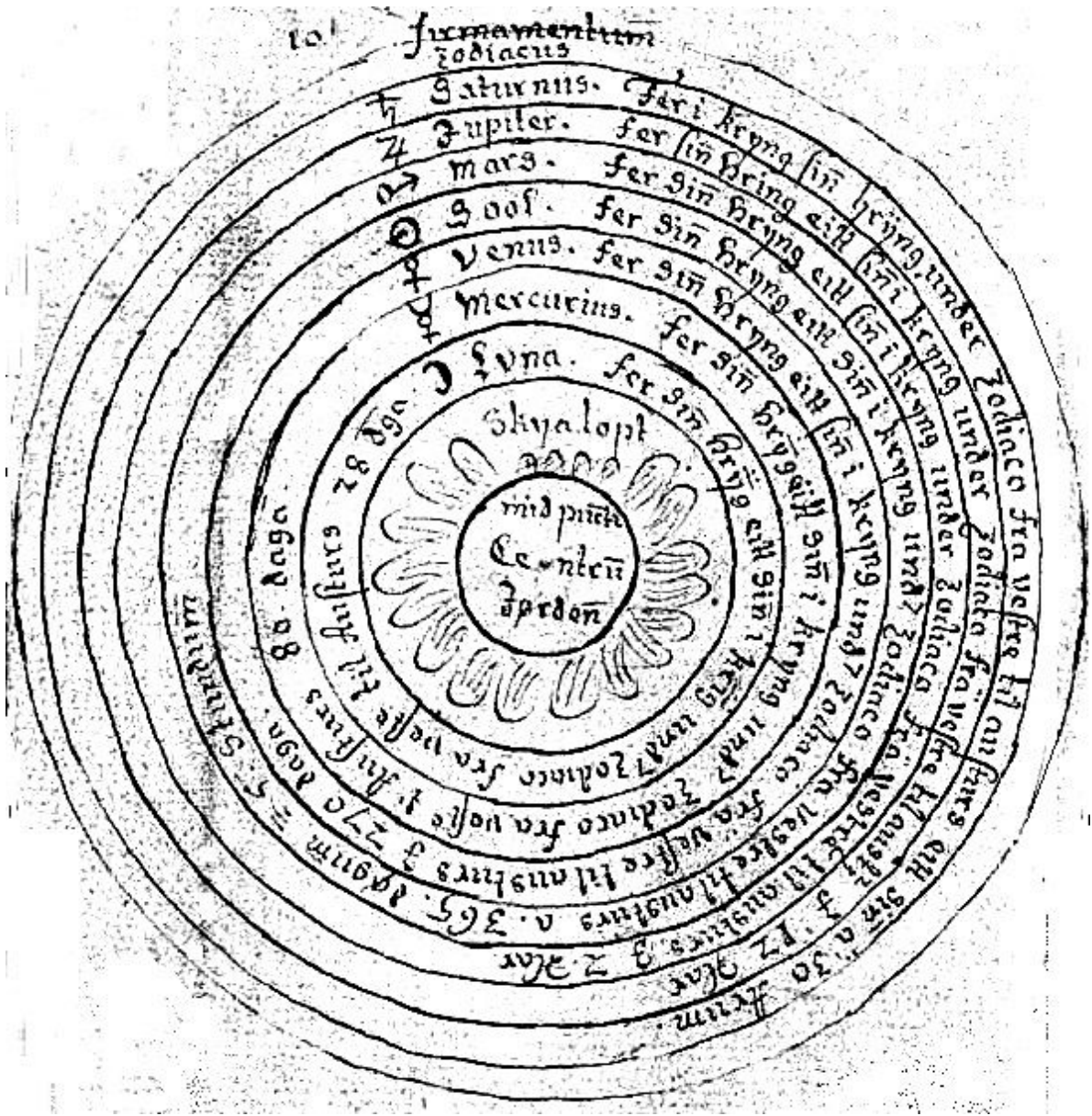

**Figure 1 - A geocentric world view, like in the Grosseteste's cosmology. Note the firmament above the sphere of the Zodiac. This image is adapted from an Icelandic manuscript, now in the care of the Magnusson Institute in Iceland. Courtesy: Wikipedia.**